\documentclass[longauth]{aa}
\usepackage{graphicx}

\usepackage{txfonts}
\usepackage[breaklinks, colorlinks, citecolor=blue, linkcolor=blue]{hyperref}
\usepackage{natbib}
\bibpunct{(}{)}{;}{a}{}{,} 
\usepackage{xspace}
\usepackage{orcidlink}

\newcommand\tess{{TESS}\xspace}

\newcommand\kepler{{\it Kepler}\xspace}

\newcommand{\eref}[1]{Equation~(\ref{#1})}
\newcommand{\fref}[1]{Fig.~\ref{#1}}
\newcommand{\tref}[1]{Table~\ref{#1}}
\newcommand{\sref}[1]{Section~\ref{#1}}

\newcommand{\vsini}{$v \sin i_\star$\xspace}
\newcommand{\teff}{$T_{\rm eff}$\xspace}
\newcommand{\feh}{$\rm [Fe/H]$\xspace}
\newcommand{\logg}{$\log g$\xspace}

\newcommand{\imp}{$0.904^{+0.005}_{-0.007}$\xspace}

\newcommand{\rpbJ}{$1.72 \pm 0.05$~R$_{\rm J}$\xspace}

\newcommand{\mpbJ}{$0.57 \pm 0.02$~M$_{\rm J}$\xspace}
\newcommand{\rhob}{$0.138 \pm 0.013$~g~cm$^{-3}$\xspace}

\newcommand{\lam}{$184\pm3$$^\circ$\xspace}
\newcommand{\lamshadow}{$189\pm8$$^\circ$\xspace}

\newcommand{\psib}{$101^{+5}_{-9}$$^\circ$\xspace}
\newcommand{\psibtwo}{$117^{+6}_{-8}$$^\circ$\xspace}
\newcommand{\psibrot}{$104\pm2$$^\circ$\xspace}
\newcommand{\psibtworot}{$121^{+2}_{-3}$$^\circ$\xspace}

\usepackage{booktabs}
\usepackage{threeparttable}
\usepackage{soul}

\usepackage[normalem]{ulem}

\begin{document}

\title{A puffy polar planet}
   \titlerunning{The obliquity of the TOI-640 system}
   \subtitle{The low density, hot Jupiter TOI-640~b is on a polar orbit\thanks{Based on observations made with the ESO-3.6~m telescope at La Silla Observatory under programme 106.21TJ.001.} }

   \author{
   Emil~Knudstrup\inst{\ref{sac}}\orcidlink{0000-0001-7880-594X} \and
    Simon~H.~Albrecht\inst{\ref{sac}}\orcidlink{0000-0003-1762-8235} \and
    Davide~Gandolfi\inst{\ref{torino}}\orcidlink{0000-0001-8627-9628} \and
    Marcus~L.~Marcussen\inst{\ref{sac}}\orcidlink{0000-0003-2173-0689} \and
    Elisa~Goffo\inst{\ref{torino},\ref{tls}} \and
    Luisa~M.~Serrano\inst{\ref{torino}} \and
    Fei~Dai\inst{\ref{geo},\ref{caltech}} \orcidlink{0000-0002-8958-0683}\and
    Seth~Redfield\inst{\ref{wesley}} \and
    Teruyuki~Hirano\inst{\ref{bio},\ref{tokyo}} \and
    Szilárd~Csizmadia\inst{\ref{dlr}} \and
    William~D.~Cochran\inst{\ref{mcd}}\orcidlink{0000-0001-9662-3496} \and
    Hans~J.~Deeg\inst{\ref{iac},\ref{ull}} \and
    Malcolm~Fridlund\inst{\ref{leiden},\ref{chalm}} \and
    Kristine~W.~F.~Lam\inst{\ref{dlr}}\orcidlink{0000-0002-9910-6088} \and
    John~H.~Livingston\inst{\ref{bio},\ref{tokyo},\ref{sokendai}}\orcidlink{0000-0002-4881-3620}\and
    Rafael~Luque\inst{\ref{chic}}\orcidlink{0000-0002-4671-2957} \and
    Norio~Narita\inst{\ref{koma},\ref{bio},\ref{iac}}\orcidlink{0000-0001-8511-2981} \and
    Enric~Palle\inst{\ref{iac},\ref{ull}} \and
    Carina~M.~Persson\inst{\ref{chalm}}\orcidlink{0000-0003-1257-5146} \and
    Vincent~Van~Eylen\inst{\ref{ucl}}
  }

   \authorrunning{Knudstrup et~al.}
   \institute{
   Stellar Astrophysics Centre, Department of Physics and Astronomy, Aarhus University, Ny Munkegade 120, DK-8000 Aarhus C, Denmark \email{emil@phys.au.dk}\label{sac} \and
   Dipartimento di Fisica, Università degli Studi di Torino, via Pietro Giuria 1, I-10125, Torino, Italy \label{torino} \and
   Thüringer Landessternwarte Tautenburg, Sternwarte 5, 07778 Tautenburg, Germany \label{tls} \and
   Division of Geological and Planetary Sciences, 1200 E California Blvd, Pasadena, CA, 91125, USA \label{geo} \and
   Department of Astronomy, California Institute of Technology, Pasadena, CA 91125, USA \label{caltech} \and
   Astronomy Department and Van Vleck Observatory, Wesleyan University, Middletown, CT 06459, USA \label{wesley} \and
   Astrobiology Center, 2-21-1 Osawa, Mitaka, Tokyo 181-8588, Japan \label{bio} \and
   National Astronomical Observatory of Japan, 2-21-1 Osawa, Mitaka, Tokyo 181-8588, Japan \label{tokyo} \and
   Institute of Planetary Research, German Aerospace Center (DLR), Rutherfordstrasse 2, 12489 Berlin, Germany \label{dlr} \and
   Center for Planetary Systems Habitability and McDonald Observatory, The University of Texas at Austin, Austin Texas USA 78712 \label{mcd} \and
   Instituto de Astrofísica de Canarias (IAC), 38200 La Laguna, Tenerife, Spain \label{iac} \and
   Departamento de Astrofísica, Universidad de La Laguna (ULL), 38206 La Laguna, Tenerife, Spain \label{ull} \and
   Leiden Observatory, P.O. Box 9513, NL-2300 RA Leiden, The Netherlands \label{leiden} \and
   Department of Space, Earth and Environment, Chalmers University of Technology, Onsala Space Observatory, SE-439 92 Onsala, Sweden \label{chalm} \and
   Department of Astronomy, The Graduate University for Advanced Studies (SOKENDAI), 2-21-1 Osawa, Mitaka, Tokyo, Japan \label{sokendai} \and
   Department of Astronomy \& Astrophysics, University of Chicago, Chicago, IL 60637, USA \label{chic} \and
   Komaba Institute for Science, The University of Tokyo, 3-8-1 Komaba, Meguro, Tokyo 153-8902, Japan \label{koma} \and
    Mullard Space Science Laboratory, University College London, Holmbury St Mary, Dorking, Surrey RH5 6NT, UK \label{ucl}
   }

   \date{Received ...; accepted ...}

  \abstract{
    TOI-640~b is a hot, puffy Jupiter with a mass of \mpbJ and radius of \rpbJ, orbiting a slightly evolved F-type star with a separation of $6.31^{+0.09}_{-0.07}$~R$_\star$. Through spectroscopic in-transit observations made with the HARPS spectrograph, we measured the Rossiter-McLaughlin effect, analysing both in-transit radial velocities and the distortion of the stellar spectral lines. From these observations, we find the host star to have a projected obliquity of $\lambda=$\lam. From the TESS light curve, we measured the stellar rotation period, allowing us to determine the stellar inclination, $i_\star=23^{+3\circ}_{-2}$, meaning we are viewing the star pole-on. Combining this with the orbital inclination allowed us to calculate the host star obliquity, $\psi=$ \psibrot. TOI-640~b joins a group of planets orbiting over stellar poles within the range $80^\circ-125^\circ$. The origin of this orbital configuration is not well understood.  
  }

   \keywords{techniques: radial velocities -- 
   techniques: photometric --
   planets and satellites: gaseous planets --
   planet-star interactions
                 }
   \maketitle
 
\section{Introduction}

Before 1992, the only planetary system we knew of was the Solar System. The neat and ordered structure we see in the Solar System therefore formed the architectural drawing for planetary formation and evolution. However, with the detection of the first exoplanet, it immediately became clear that this schematic does not apply to all systems. For instance, the very first exoplanet discovered is orbiting a pulsar \citep{Wolszczan1992}, the first exoplanet around a Sun-like star is a Jupiter-sized planet on a $\sim$4~d orbit \citep{Mayor1995}, and some systems harbour planets on wildly eccentric orbits \citep[e.g.][]{Cochran2008}. The type of host stars, the orbital separations, and eccentricities are just some of the parameters indicating how different exoplanet
systems can be from the Solar System.

Another parameter is the angle between the stellar spin axis of the host and the orbital axis of the planet, the spin-orbit angle, or the obliquity $\psi$.\footnote{We note that here we are discussing the obliquity of the host star and not the planet. In this article, we use the terms obliquity and spin-orbit angle interchangeably.}  At  $7.155\pm 0.002^\circ$ \citep{Beck2005}, the obliquity of the Solar System is relatively low.  In contrast, in exoplanet systems, measurements of $\psi$, or its projection on the sky $\lambda$, or the difference along the line of sight between orbital and stellar spin, display a large variety of values. The configurations range from well aligned to (moderately) misaligned, and there are even retrograde systems \citep[see e.g. the review by][ and references therein]{Albrecht2022}. There is also a curious trend reported by \cite{Albrecht2021}; systems for which $\psi$ has been measured are either consistent with good alignment or the planets orbit over the stellar poles. This preponderance of perpendicular planets was not evident from $\lambda$ measurements alone, as without additional knowledge, meaningful inferences about $\psi$  cannot be drawn from $\lambda$ measurements \citep{Fabrycky2009}.

Here we aim to measure the host star obliquity in the TOI-640 system discovered and characterised by \citet{Rodriguez2021}. To this end, we make use of the Rossiter-McLaughlin (RM) effect, an apparent distortion of the stellar line shapes caused by a transiting body blocking part of the rotating stellar disk. The RM effect allows us to measure the sky-projected obliquity, $\lambda$. To measure the stellar inclination, we use light curves from the Transiting Exoplanet Survey Satellite \citep[TESS;][]{Ricker2015}. Together with knowledge of the orbital inclination, we can infer the spin-orbit angle of our target system.  

The paper is organised as follows. In \sref{sec:obs}, we present the observations, both photometric and spectroscopic. \sref{sec:obl} presents the determination of the obliquity of the  host star. Our new radial velocities (RVs) and photometry allow us to also update a number of other system parameters. We discuss these together with our result on the spin-orbit angle in \sref{sec:disc} before giving our conclusions in \sref{sec:con}.

\section{Observations}\label{sec:obs}

\subsection{\tess photometry}
\tref{tab:pars}  lists a selection of parameters determined by \citet{Rodriguez2021}. These authors presented \tess data of TOI-640 from Sectors 6 and 7 taken with a cadence of 30~min. Additional \tess photometry has become available since then, as the system was observed again in Sectors 33 and 34. This time the system was observed with a cadence of 2~min. \fref{fig:allTESS}  displays the \tess data from all four sectors.

\begin{table}[]
    \centering
    \caption{{\bf Literature system parameters.} }
  \begin{threeparttable} 
    \begin{tabular}{l l c}
    \toprule
    Parameter   & Value \\
    \midrule
    Stellar mass,  $M_\star$  (M$_\odot$) & $1.536^{+0.069}_{-0.076}$ \\
    Stellar radius,  $R_\star$  (R$_\odot$) & $2.082^{+0.064}_{-0.058}$ \\
    Effective temperature, \teff (K) & $6460^{+130}_{-150}$ \\
    Surface gravity, \logg (dex) & $3.987^{+0.030}_{-0.036}$ \\
    Metallicity (dex), \feh  & $0.072^{+0.085}_{-0.076}$ \\
    Age, $\tau$ (Gyr) & $1.99^{+0.55}_{-0.40}$ \\
    Proj. rotational velocity, \vsini (km~s$^{-1}$) & $6.1 \pm 0.5$ \\
    Macroturbulence, $v_{\rm mac}$ (km~s$^{-1}$) & $6.32 \pm 1.37$ \\
    \midrule
    Orbital period, $P$ (days) & $5.0037775(48)$ \\
    Planet-to-star radius ratio, $R_\mathrm{p}/R_\star$& $0.08738^{+0.00091}_{-0.00086}$ \\ 
    Semi-major axis / star radius, $a/R_\star$ & $6.82^{+0.22}_{-0.24}$ \\ 
    Velocity semi-amplitude, $K$ (m s$^{-1}$) & $78\pm14$ \\ 
    Impact parameter,   $b$ & $0.8763^{+0.0063}_{-0.0067}$ \\
    Eccentricity, $e$ & $0.050^{+0.054}_{-0.035}$\\
      \midrule
    Planet radius,  $R_\mathrm{P}$ (R$_{\rm J}$) & $1.771^{+0.060}_{-0.056}$ \\
    Planet mass, $M_\mathrm{P}$ (M$_{\rm J}$) & $0.88 \pm 0.16$ \\
     \bottomrule
    \end{tabular}
 \begin{tablenotes}
    \item Selected stellar, orbital, and planetary parameters from \citet{Rodriguez2021}.
\end{tablenotes}
  \end{threeparttable} 
    \label{tab:pars}
\end{table}

We downloaded and reduced the \tess data utilising the python package \texttt{lightkurve} \citep{lightkurve}. First, we corrected for noise induced by the motion of the spacecraft and removed scattered light using the \texttt{RegresssionCorrector} routine. The result is shown in the top panel of \fref{fig:allTESS}. To exclude outliers, we then (temporarily) removed the transits from the planet using the best-fitting transit parameters, which were determined by fitting the light curve iteratively. The resulting light curves are shown in the middle panel of \fref{fig:allTESS}, where we also overplotted a Savitzky-Golay \citep{savitzky1964} filter to (again temporarily) smooth the light curve. Points more than 5$\sigma$ away from the smoothed light curves were rejected (19 out of 36,918 points were removed). The bottom panel of  \fref{fig:allTESS} displays  the unfiltered light curves with outliers excluded, but with the transits re-injected. This is the light curve we use in the analysis for determining the projected spin-orbit angle in \sref{sec:RM}, whereas we use the light curve with the transits removed for determining the rotation period in \sref{sec:inclPhotRot}. 

\subsection{HARPS spectroscopy}

\begin{figure}
    \centering
    \includegraphics[width=\columnwidth]{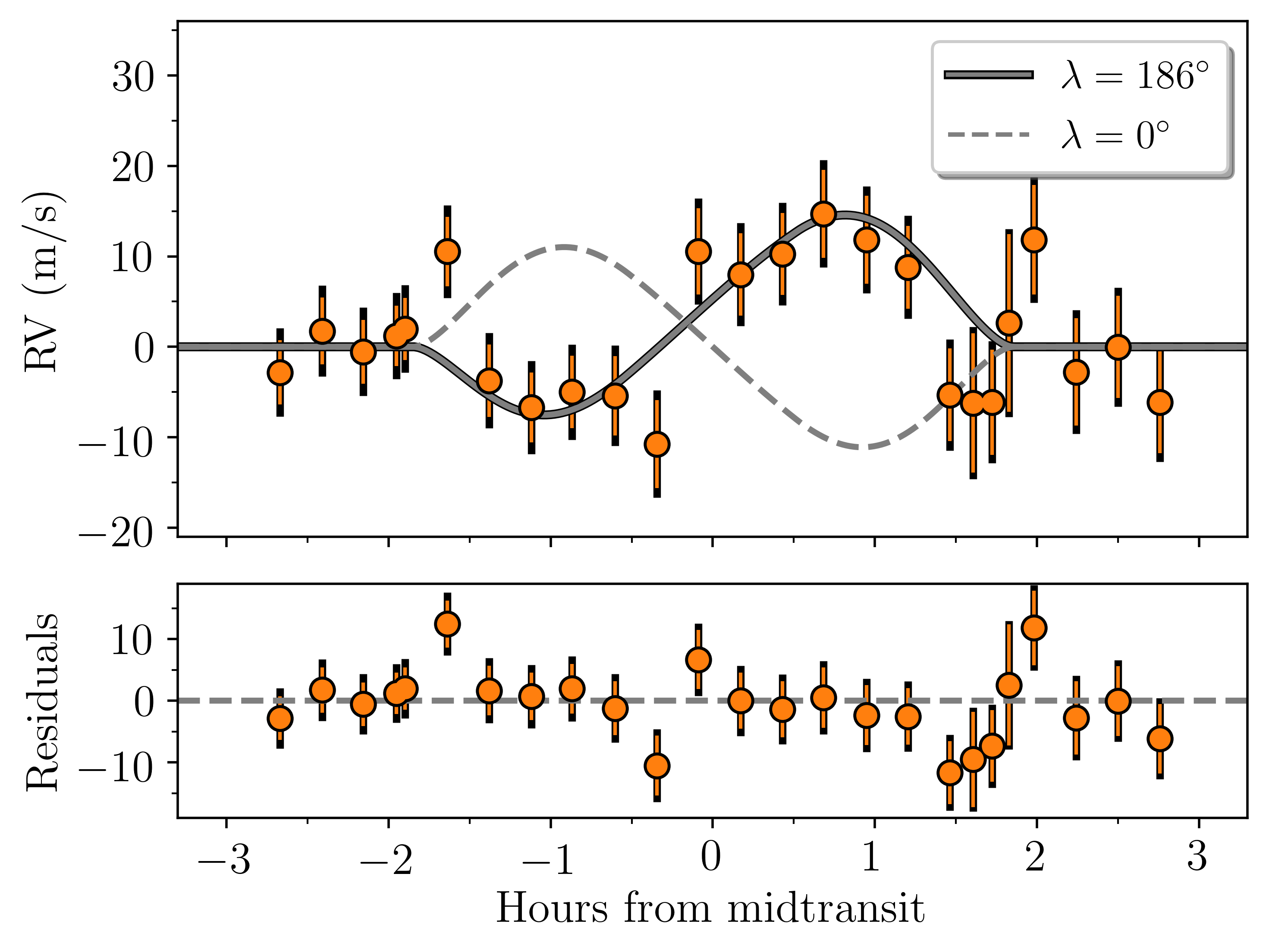}
    \caption{{\bf Rossiter-McLaughlin effect as seen from} HARPS RVs centred around the mid-transit time after subtracting the Keplerian motion induced by the planet. The grey line shows the RM effect with the best-fitting (retrograde) model as the solid line and an aligned model as the dashed line. The error bars include the jitter term from our MCMC added in quadrature, shown as the black extension. }
    \label{fig:rmrv}
\end{figure}

To measure the RM effect, we observed a transit of TOI-640~b that occurred during the night 2022 February 26 UT using the High-Accuracy Radial Velocity Planetary Searcher \citep[HARPS;][]{mayor2003} as part of our observing programme 106.21TJ.001 (PI: Gandolfi). HARPS is mounted at the European Southern Observatory (ESO)-3.6~m telescope at La Silla, Chile. We obtained 22 spectra on the transit night of which 4 were acquired before the beginning of ingress, 14 during transit, and 4 after egress. The average exposure time was 900~s and the median signal-to-noise ratio (S/N) per pixel at 550~nm was 45. We continued to monitor the system using HARPS until 2022 November 23 UT covering a total time span of 271 days with an additional 40 radial velocities. 
For these monitoring observations, the exposure times varied between 1200~s and 1500~s depending on sky conditions, and the median S/N was 50. The extracted RVs, their associated errors, and photometric mid-times are shown in \fref{fig:rv_grand} and listed in \tref{tab:trvs} and \tref{tab:rvs}. \fref{fig:rmrv}  shows the HARPS RVs centred on the mid-transit time, focusing in on the RM effect.

\section{Stellar obliquity}\label{sec:obl}

In this section, we first conduct a joint fit of both the photometric and spectroscopic data to measure the projected spin-orbit angle of TOI-640 (\sref{sec:RM}). We then determine the stellar inclination along the line of sight making use of \tess photometry (\sref{sec:inclPhotRot}) and through the use of an empirical relation (\sref{sec:inclLouden}). Together with the orbital inclination, we can then determine the stellar obliquity (\sref{sec:psi}). Our results for $\psi$ as well as other system parameters are then discussed in \sref{sec:disc}.

\begin{figure*}
    \centering
    \includegraphics[width=\textwidth]{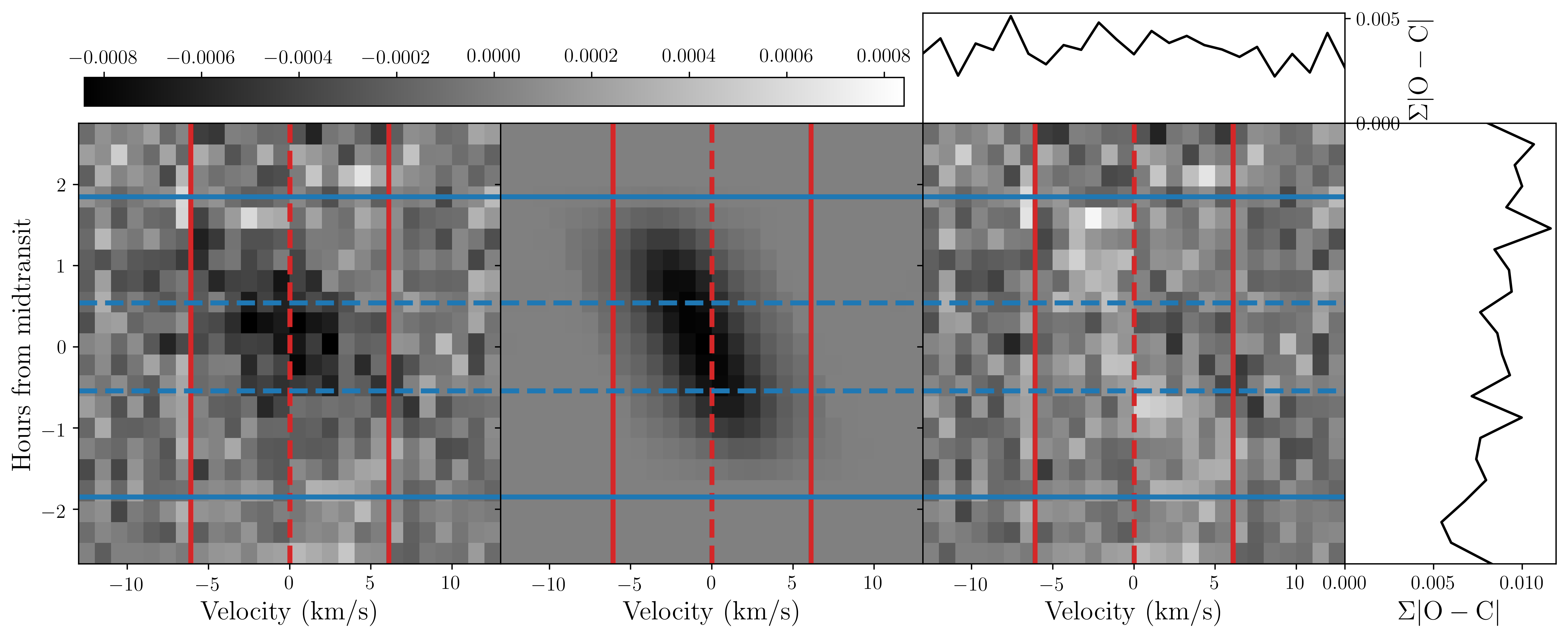}
    \caption{{\bf Planetary shadow.} {\it Left:} Distortion of the stellar absorption lines as seen for our observations with HARPS taken on the transit night. The vertical solid lines denote $\pm v \sin i_\star$, and the vertical dashed line is at $v=0$~km~s$^{-1}$. The horizontal dashed lines mark the points of second and third contact, which is when the planet is found completely within the stellar disk, and the solid lines denote the first and fourth contact points, where the planetary and stellar disk start to overlap. {\it Middle:} Best-fitting model of the distortion. {\it Right:} Residuals from subtracting the best-fitting model from the data. The horizontal bar on top shows the strength of the signal. }
    \label{fig:shadow}
\end{figure*}

\subsection{Projected obliquity from the 
Rossiter-McLaughlin effect}\label{sec:RM}

Spectrally resolved stellar lines observed during transits (or eclipses) will display distortions; this is known as the RM effect. For example, in a system where the projections of the stellar spin axis and the orbital axis of the planet are aligned (low projected obliquity), a transiting planet would first hide sections of the approaching stellar surface. A distortion of the lines with a negative velocity (relative to the current RV of the star) will appear. During later phases of the transit, further 
areas of the stellar surface with positive radial velocities (redshifted) will be hidden from view. This distortion can also be sensed as anomalous RVs during transits. In this case, first positive and later negative RVs are observed. However, if the spin-orbit angle is larger than $90^\circ$ (a retrograde configuration) then the time evolution of the distortion and RVs is reversed. From a glance at \fref{fig:rmrv} and \fref{fig:shadow}, it appears as if the orbit of TOI-640\,b is indeed retrograde.

Our approach to a quantitative analysis of the RM effect is similar to the analyses performed in \citet{Knudstrup2022a,Knudstrup2022b}, and we briefly summarise it here. Also,  the following procedure is included in the python package \st{tracit}\footnote{\url{https://tracit.readthedocs.io/en/latest/}}, which was used in these latter two publications. 

We performed a joint fit of the photometric and spectroscopic data. Specifically, we performed two different analyses of the spectroscopic data. First, we analysed the anomalous RVs obtained during the transit. We then performed a second analysis where we do not use the RVs during transit but the underlying distortions of the line shapes, or the so-called planet shadow. We did this to check for consistency between the different measurement approaches \citep{Albrecht2007}. In both cases, we also made use of the orbital RV measurements and the TESS photometry and we applied the same priors. 

When determining $\lambda$ through the anomalous RVs, we used the RVs obtained from the HARPS Data Reduction Software \citep[DRS;][]{Lovis2007}. When analysing the planetary shadow, we used the cross-correlation functions (CCFs) obtained from the DRS as a proxy for the stellar absorption lines.
With HARPS' resolution of $R=120\,000,$ the point spread function (PSF) has a full width at half maximum (FWHM) of $2.5$~km\,s$^{-1}$ or an  equivalent dispersion of $\sigma\approx1$~km\,s$^{-1}$. The CCFs delivered by the HARPS DRS are over-sampled with a datum every $0.25$~km\,s$^{-1}$. To account for this, we interpolated the CCFs onto a grid with a resolution of $1$~km~s$^{-1}$. This is the same approach as taken in \cite{Knudstrup2022a} and similar to the approach taken by \citet{Cegla2016} for instance, where every fourth datum in the grid is sampled. 

The relevant parameters in both approaches are the orbital period $P$, mid-transit time $T_0$, planet-to-star radius ratio $R_{\rm p}/R_\star$, scaled semi-major axis $a/R_\star$, orbital inclination $i$, RV semi-amplitude $K$, orbital eccentricity $e$, argument of periastron $\omega$, projected stellar obliquity $\lambda$, projected stellar rotation speed $v \sin i_\star$, macro-turbulence $\zeta$, micro-turbulence $\xi$, systemic velocity $\gamma$, and two sets of separate pairs of quadratic limb-darkening coefficients $c_1,c_2$, for \tess and HARPS. 

We applied Gaussian priors on \vsini (from \tref{tab:pars}), as well as $\zeta$ and $\xi$ estimated from the relations in \citet{Doyle2014} and \citet{Bruntt2010}, respectively, using the parameters in \tref{tab:pars}. The Gaussian priors for the limb-darkening coefficients were obtained from the tables by \citet{Claret2013} and \citet{Claret2018} for HARPS and \tess, respectively. Uncertainties of $0.1$ were assumed. Uniform priors were applied for all other parameters.

To model the RM effect for the RVs, we used the code by \citet{Hirano2011}, while we used the formulation in \citet{Albrecht2007,Albrecht2013} to model the planetary shadow. We modelled the \tess data using the {\tt batman} package \citep{Kreidberg2015}. This was done with the inclusion of Gaussian process (GP) regression ---utilising the library {\tt celerite} \citep{celerite}--- to characterise the photometric noise (stellar and instrumental). For our GP, we used a Matérn-3/2 kernel, which is characterised by two hyperparameters; the amplitude, $A$, and the timescale, $\tau$. We sampled the posterior distribution for the parameters through Markov Chain Monte Carlo (MCMC) sampling using the code {\tt emcee} \citep{emcee}. In our MCMC, we stepped in $\sqrt{e} \cos \omega$ and $\sqrt{e} \sin \omega$ as opposed to stepping in $e$ and $\omega$ directly. For the limb-darkening parameters, we stepped in the sum of the coefficients while keeping the difference fixed. Furthermore, we stepped in $\cos i$ instead of $i$, allowing us to apply a flat prior assuming an isotropic spin distribution a priori. All stepping parameters and priors are listed in \tref{tab:results} and \tref{tab:ld}. Our likelihood is defined as
\begin{equation}
    \log \mathcal{L} = -0.5 \sum_{i=1}^{N} \left [ \frac{(O_i - C_i)^2}{\sigma_i^2} + \log 2 \pi \sigma_i^2 \right]+ \sum_{j=1}^{M}  \mathcal{P}_{j}\, ,
\end{equation}
where $N$ indicates the total number of data points from photometry and RVs. $C_i$ represents the model corresponding to the observed data point $O_i$, and $\mathcal{P}_j$ is the prior on the $j$th parameter.

Finally, before starting the joint spectroscopic photometric MCMC runs, we performed simple minimisations on each of the three data types. We then added `jitter' terms 
in quadrature to the respective uncertainties until reduced $\chi^2$ of 1 were obtained. This was done in an attempt to ensure proper weighting between spectroscopic and photometric data.

\begin{figure}
    \centering
    \includegraphics[width=\columnwidth]{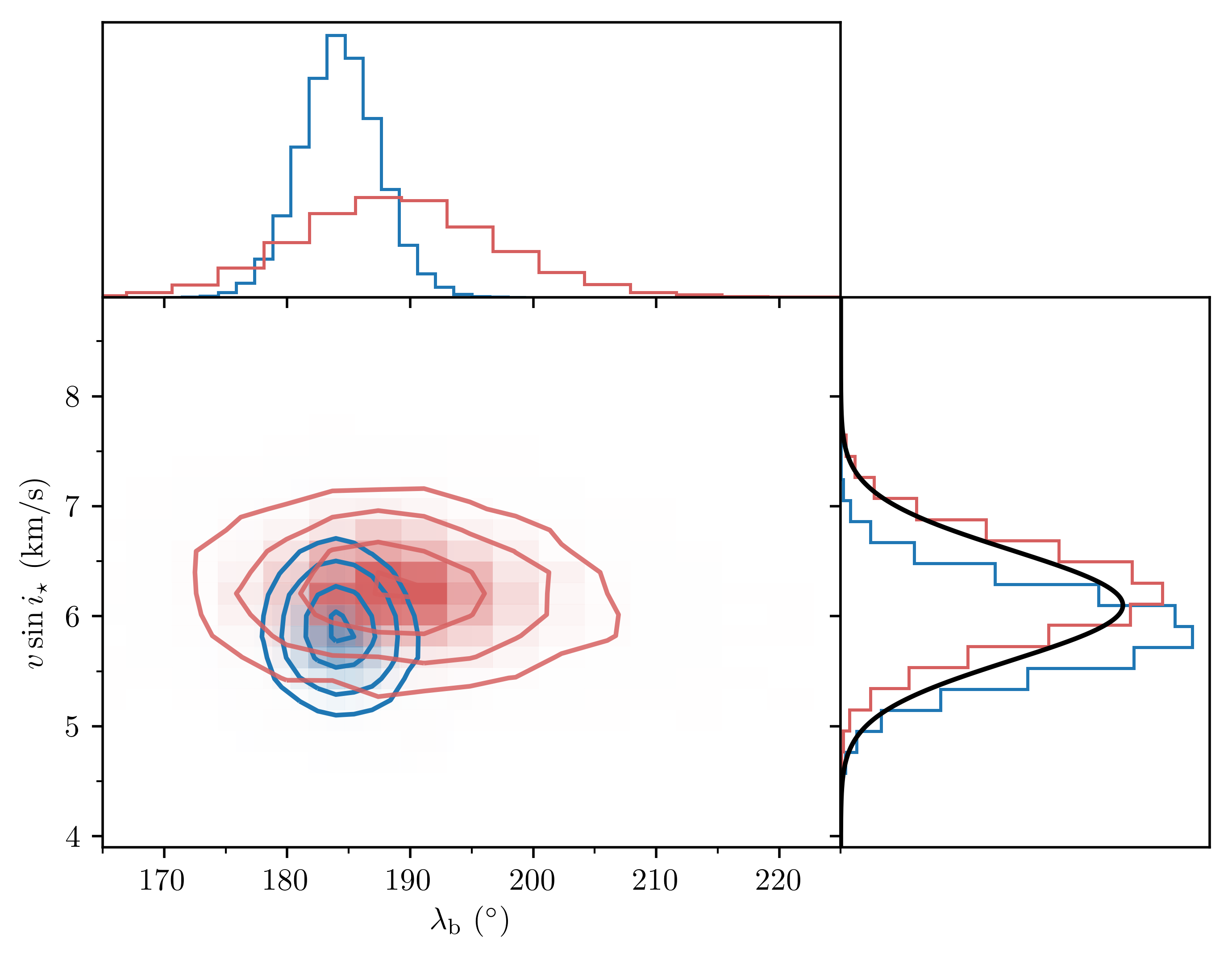}
    \caption{{\bf 2D histograms.} The correlation between $\lambda$ and \vsini from our MCMCs. Blue shows the results from our RV run, while red is from the shadow run.}
    \label{fig:scorner}
\end{figure}

The observed RVs and the best-fitting model are shown in \fref{fig:rmrv}, and the results are presented in the fourth column of \tref{tab:results}. The observed shadow and best-fitting model can be seen in \fref{fig:shadow} with the results presented in the fifth column of \tref{tab:results}. For both runs, RVs and shadow, we present the nuisance parameters in \tref{tab:ld}. We show the correlation plot for \vsini and $\lambda$ for both runs in \fref{fig:scorner}. An extended correlation plot for more parameters can be found in \fref{fig:corner_app}.

The amplitude of the RM signal (relative to the noise) seen in \fref{fig:rmrv} and \fref{fig:shadow} is modest; nevertheless the uncertainties in our $\lambda$ measurements are comparably low. This is because the large impact parameter of \imp acts as a lever. Even a small change in $\lambda$ leads to a transit chord passing over stellar surface areas with substantially different rotational RVs. 

The $\lambda$ measurement from the analysis of the RVs comes out to \lam, and has a lower uncertainty than the measurement from the planet shadow \lamshadow. This is because the overall line width ---mainly governed by $v\sin i_\star$ relative to the width of the distortion--- is only modest (\fref{fig:shadow}). A larger $v\sin i_\star$ value would have reversed the situation as it would have lead to a more localised distortion (planet shadow) in the lines \citep{Albrecht2022}. At the same time, a larger $v\sin i_\star$ leads to a larger RV uncertainty. These two advantages of analysing the line distortions relative to analysing the RVs vanish for lower $v\sin i_\star$. In addition, the shadow measurement requires alignment and normalisation of the CCFs \citep{Knudstrup2022a}. This takes away some of the predictive power from the CCFs as these additional `hyper parameters' (not to be confused with the GP hyper parameters) have to be determined \citep{Albrecht2013b}. This process is similar to the `Hyperplane Least Squares' method described and tested by \citet{Bakos2010}. Here, for the case of TOI-640, we therefore use the values obtained via the analysis of the anomalous RVs moving forward.

Up to this point, we have determined $\lambda$, the projected spin-orbit angle. Next, we determine the stellar inclination, $i_\star$, using two different methods. Together with the orbital inclination, this allows us to determine the obliquity, $\psi$.

\begin{figure}
    \centering
    \includegraphics[width=\columnwidth]{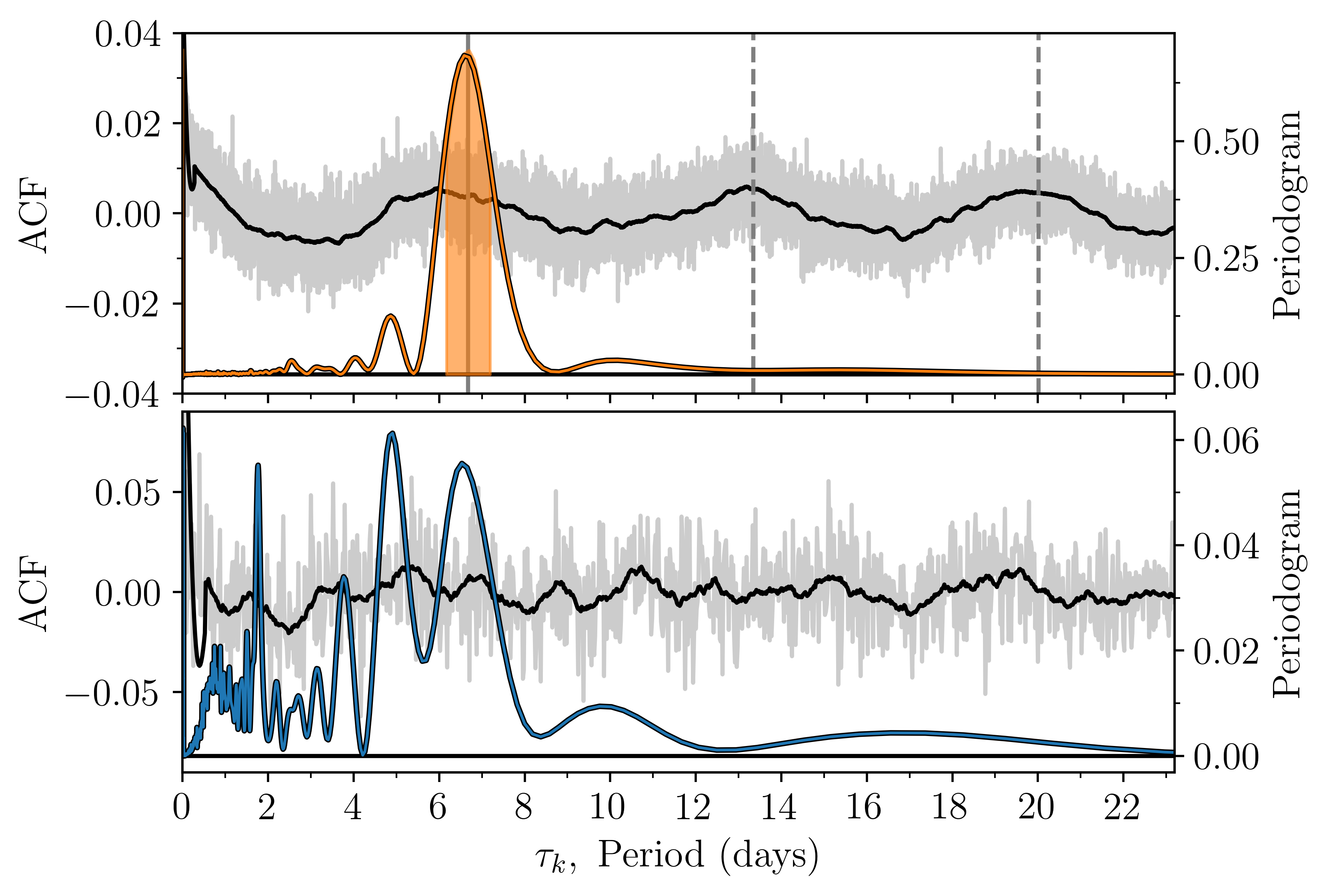}
    \caption{{\bf Autocorrelation function.} The ACFs are shown as the grey curves with a smoothed version in black. The (coloured) periodogram is calculated from the smoothed ACF. The $y$-axis on the left (right) is for the ACF (periodogram). {\it Top:} The 2~min cadence ACF with the corresponding periodogram. The vertical grey line denotes the measured rotation period, and the dashed lines are integer values of this value. Shown as the shaded area is the confidence interval for the rotation period. {\it Bottom:} The 30~min cadence ACF.}
    \label{fig:acf}
\end{figure}

\subsection{Stellar inclination from \tess photometry}\label{sec:inclPhotRot}

The starting point for our first inclination measurement is the rotation period of the star, $P_{\rm rot}$, as determined from \tess photometry in \fref{fig:allTESS}. We used the light curves in the middle, where the transits have been removed by the best-fitting transit model. We measured the rotation period using the autocorrelation method \citep[e.g.][]{McQuillan2013}. We do this by calculating the autocorrelation function (ACF) for the 2~min cadence and 30~min cadence separately. We then smoothed the ACF using a Savitzky-Golay filter from which we calculated the Generalised Lomb-Scargle \citep[GLS;][]{Lomb1976,Scargle1982} periodogram.

While we can clearly identify a single peak (at around 6.7 days) in the periodogram for the 2~min cadence case, we also see some additional features most likely associated with momentum dumps of the spacecraft causing `jumps' in the light curves. An example of how these jumps affect the ACF and periodogram is illustrated in the third column of Fig.~2 in \citet{McQuillan2013}. We therefore applied a Savitzky-Golay filter to the light curves to remove these jumps. The resulting ACFs and periodograms are shown in \fref{fig:acf}. Evidently, the rotation is detected much more clearly for the 2~min cadence case, but it is also apparent in the 30~min cadence. We therefore proceeded with the results from the 2~min cadence periodogram.

We fitted a Gaussian to the peak in the periodogram to get an estimate of the period and the uncertainty. From this, we got a rotation period of $P_{\rm rot} = 6.7 \pm 0.6$~d, which we can use with $R_\star=2.082\pm0.061$~R$_\odot$ to calculate the stellar inclination from
\begin{equation}
    \sin i_\star = \frac{P_{\rm rot}v \sin i_\star}{2 \pi R_\star} \, .
    \label{eq:sini}
\end{equation}
We followed the approach in \citet{Masuda2020} to properly calculate $i_\star$ from \eref{eq:sini}, meaning we accounted for the fact that $v$ and \vsini are not independent. From this, we get a rotation speed at the equator of $v=14\pm2$~km~s$^{-1}$ and subsequently a stellar inclination of $i_\star=23^{+3}_{-2}$$^\circ$.

\begin{figure*}
    \centering
    \includegraphics[width=\textwidth]{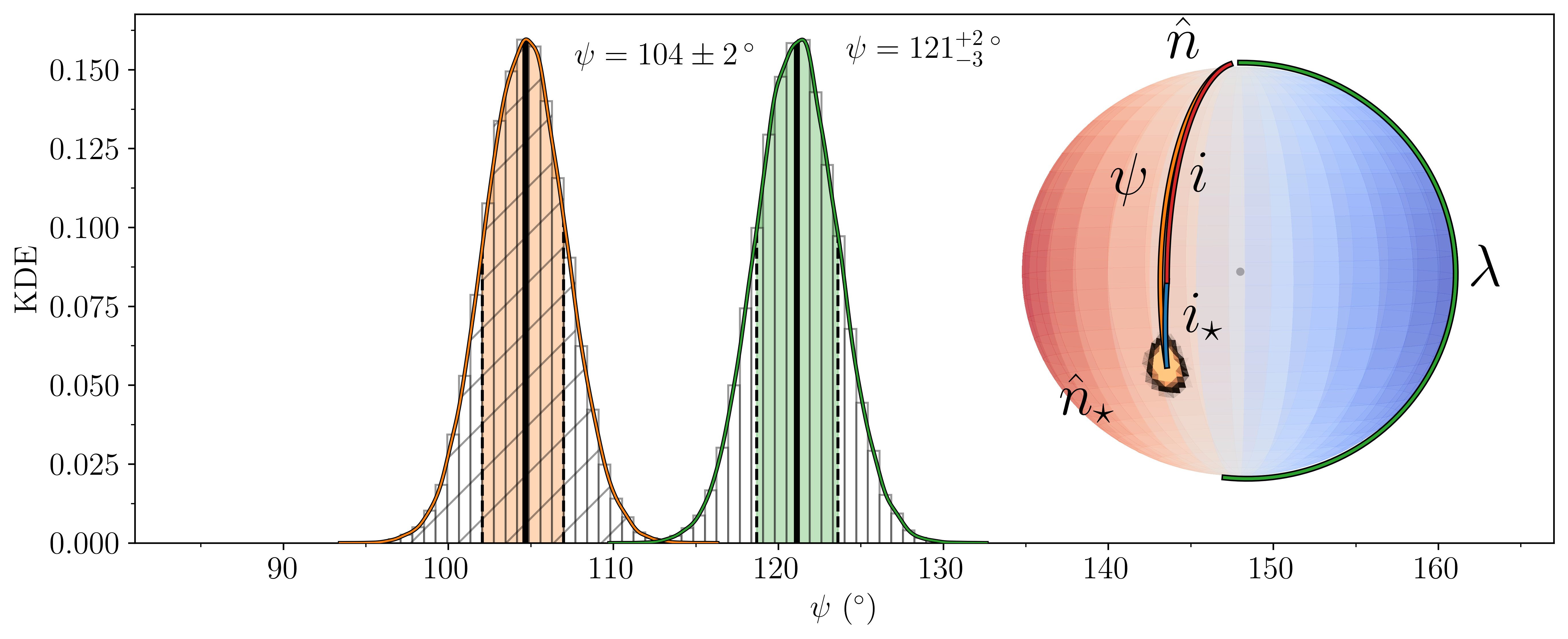}
    \caption{{\bf Obliquity distributions.} The histograms are the distributions for $\psi$ calculated from the rotation period in \sref{sec:inclPhotRot} with the KDE overplotted as the solid lines. The coloured areas are the confidence intervals with the medians shown as the black lines. The orange distribution corresponds to using the orbital inclination, $i$, directly from our posteriors and the green is $i-180^\circ$. On the sphere, we only show the ``orange" distribution for $\psi$ as the heatmap. We do this for a fixed value of $i$ (in terms of how the sphere is oriented). The sphere is colour coded according to the perceived movement of the stellar disk with the approaching (receding) side in blue (red) for an observer along the line of sight, which is denoted by the grey arrow. }
    \label{fig:psi}
\end{figure*}

\subsection{Stellar inclination from the empirical relation}\label{sec:inclLouden}

In the second approach, we used the results from \citet{Louden2021}, who investigated the obliquities of \kepler planets around hot stars. This required that the authors determine the $v$ distribution of a sample of control stars without detected transiting planets. From that sample, they obtained a relation between the mean rotation velocity of a star as a function of the effective temperature. We can use this relation with the \teff from \tref{tab:pars} to calculate $i_\star$. We obtain $i_\star=20^{+6}_{-9}$~deg. 

Using this relation from \citet{Louden2021} would not be appropriate in the case of tidal spin-up in which tides raised on the star by the planet change the rotation of the star. This effect has been suggested to take place in some hot Jupiter systems \citep[e.g.][]{Brown2014,Maxted2015,Yee+2022}. How effectively the planet can raise tides on the star is especially dependent on the separation, but also on the mass ratio \citep[see e.g. ][]{Zahn1977}. A useful metric to assess whether tides are effectively raised is given by $(M_{\rm p}/M_\star) ( R_\star/a  )^3$ which for TOI-640~b is $1.4 \times 10^{-6}$, meaning that tides should not play an important role and suggesting that the use of the relation from \citet{Louden2021} is warranted. For our final result for the spin-orbit angle, we use the stellar inclination measurement from the previous section and simply note here that the two inclination measurements from this and the above section are consistent.

\subsection{Stellar obliquity}\label{sec:psi}

As we now have values for $i_\star$, $i$, and $\lambda$, we can use
\begin{equation}
    \cos \psi = \sin i_\star \sin i \cos \lambda + \cos i_\star \cos i \, 
    \label{eq:psi}
\end{equation}
to calculate the spin-orbit angle, $\psi$. Here we used our distributions for $\lambda$ and $i$ from our MCMC (\tref{tab:results}), and we drew normally distributed values from $i_\star=23 \pm 2.5 ^\circ$ (determined above in Section~\ref{sec:inclPhotRot}) for each of these draws. There are two solutions. This is because our observations cannot distinguish between $i_\star$ and $180^\circ-i_\star$; they can also not distinguish between ($i, \lambda$) and $(180^\circ - i, -\lambda$). In the case of an exactly edge-on orbit ($i= 90^\circ$), the two solutions would collapse into one.
The resulting histogram and kernel density estimation (KDE) are shown in orange and green in \fref{fig:psi}. We find \psibrot or \psibtworot for the spin-orbit angle. If the orbital inclination were exactly edge on ($i=90^\circ$), then our result for the obliquity would be $\psi=113^{+3\circ}_{-2}$.

If we had not been able to determine the stellar rotation period from the TESS light curve then we could have attempted to determine $\psi$ from the inclination measurement obtained in \sref{sec:inclLouden}. In that case, we would have found a spin-orbit angle of \psib or \psibtwo. The resulting histograms and KDEs for this approach are shown in \fref{fig:app_psi}. The results are consistent between the two inclination estimates. We report the spin-orbit angle estimated from the rotation period for the conventional orientation ($i$ not $180^{\circ}-i$) as our final result, which we find to be  $\psi$ = \psibrot .

\begin{table*}
    \centering
    \caption{{\bf MCMC results.} }
    
    \begin{threeparttable}
    \begin{tabular}{c c c c c}
\toprule 
  &   &   & RV & Shadow \\ 
Parameter & Name & Prior & \multicolumn{2}{c}{Value} \\ 
\midrule 
\multicolumn{5}{c}{Stepping parameters} \\ 
\midrule 
$P$ & Period (days) & $\mathcal{U}$ & $5.003777^{+0.000002}_{-0.000003}$ & $5.003777 \pm 0.000003$ \\ 
$T_\mathrm{0}$ & Mid-transit time (BTJD) & $\mathcal{U}$ & $1459.7413 \pm 0.0003$ & $1459.7414 \pm 0.0003$ \\ 
$R_\mathrm{p}/R_\star$ & Planet-to-star radius ratio & $\mathcal{U}$ & $0.0851^{+0.0003}_{-0.0004}$ & $0.0850^{+0.0004}_{-0.0003}$ \\ 
$a/R_\star$ & Semi-major axis to star radius ratio & $\mathcal{U}$ & $6.33^{+0.07}_{-0.06}$ & $6.32^{+0.08}_{-0.07}$ \\ 
$K$ & Velocity semi-amplitude (m s$^{-1}$) & $\mathcal{U}$ & $50.1 \pm 1.0$ & $50.0^{+1.1}_{-1.2}$ \\ 
$\cos i$ & Cosine of inclination & $\mathcal{U}$ & $0.143^{+0.002}_{-0.003}$ & $0.143^{+0.002}_{-0.003}$ \\ 
$\sqrt{e} \cos \omega$ &   & $\mathcal{U}$ & $0.00^{+0.06}_{-0.07}$ & $0.02 \pm 0.07$ \\ 
$\sqrt{e} \sin \omega$ &   & $\mathcal{U}$ & $0.06^{+0.03}_{-0.06}$ & $0.07^{+0.03}_{-0.07}$ \\ 
$\lambda$ & Projected obliquity ($^{\circ}$) & $\mathcal{U}$ & $184 \pm 3$ & $189 \pm 8$ \\ 
$v \sin i_\star$ & Projected rotational velocity (km s$^{-1}$) & $\mathcal{N}$(6.1,0.5) & $5.9 \pm 0.4$ & $6.2 \pm 0.4$ \\ 
$\zeta$ & Macro-turbulence (km s$^{-1}$) & $\mathcal{N}$(6.65,1.0) & $6.6^{+0.9}_{-0.8}$ & $7.3 \pm 0.8$ \\ 
$\xi$ & Micro-turbulence (km s$^{-1}$) & $\mathcal{N}$(1.52,1.0) & $1.7 \pm 0.8$ & $1.6^{+0.8}_{-0.9}$ \\ 
$\gamma_\mathrm{HARPS}$ & Systemic velocity HARPS (m s$^{-1}$) & $\mathcal{U}$ & $40525.4 \pm 0.7$ & $40525.6^{+0.9}_{-0.8}$ \\ 
$\sigma_\mathrm{HARPS}$ & Jitter HARPS (m s$^{-1}$) & $\mathcal{U}$ & $3.9 \pm 0.7$ & $4.2 \pm 1.0$ \\ 
\midrule 
\multicolumn{5}{c}{Derived parameters} \\ 
\midrule 
$e$ & Eccentricity & - & $<0.013$~($1\sigma$) & $<0.016$~($1\sigma$) \\ 
$\omega$ & Argument of periastron ($^\circ$) & - & $87^{+50}_{-55}$ & $74^{+35}_{-63}$ \\ 
$i$ & Inclination ($^\circ$) & - & $81.79^{+0.16}_{-0.12}$ & $81.78^{+0.18}_{-0.14}$ \\ 
$b$ & Impact parameter & - & $0.904^{+0.005}_{-0.007}$ & $0.904^{+0.006}_{-0.008}$ \\ 
$T \rm _{41}$ & Total transit duration (hours) & - & $3.696^{+0.018}_{-0.019}$ & $3.696^{+0.019}_{-0.020}$ \\ 
$T \rm _{21}$ & Time from 1st to 2nd contact (hours) & - & $1.32 \pm 0.04$ & $1.31^{+0.04}_{-0.05}$ \\ 
\midrule 
\multicolumn{5}{c}{Physical parameters \tnote{\textdagger}} \\ 
\midrule 
$T_\mathrm{eq}$\tnote{$\chi$} & Equilibrium temperature (K) & - &$ 1816 \pm 39 $ & $ 1817 \pm 39 $ \\ 
$R_\mathrm{p}$ & Planet radius (R$_{\rm J}$) & - &$ 1.72 \pm 0.05 $ & $ 1.72 \pm 0.05 $ \\ 
$M_\mathrm{p}$ & Planet mass (M$_{\rm J}$) & - &$ 0.57 \pm 0.02 $ & $ 0.57 \pm 0.02 $ \\ 
$\rho_\mathrm{p}$ & Planet density (g~cm$^{-3}$) & - &$ 0.138 \pm 0.013 $ & $ 0.138 \pm 0.013 $ \\ 
$\rho_\mathrm{p}$ & Planet density ($\rho_{\rm J}$) & - &$ 0.104 \pm 0.010 $ & $ 0.104 \pm 0.010 $ \\
\bottomrule 

  \end{tabular}
 \begin{tablenotes}
    \item The median and high posterior density at a confidence level of 0.68. $\mathcal{U}$ or $\mathcal{N}$ denotes that a uniform or a Gaussian prior, respectively, was applied during the run. Barycentric TESS Julian Date (BTJD) is defined as BJD-2457000.0, BJD being the Barycentric Julian Date.
\end{tablenotes}
  \end{threeparttable}  
  \label{tab:results}
\end{table*}

\section{Discussion}\label{sec:disc}

\subsection{Refined parameters for TOI-640~b}\label{sec:ref}

From our joint fit of the photometry and the in- and out-of-transit RVs, in addition to $\lambda$, we also provide new values for other system parameters and list them in \tref{tab:results}. Some of these new determinations differ significantly from previous determinations and we discuss these first before we discuss the implications of our obliquity measurement.

\begin{figure}
    \centering
    \includegraphics[width=\columnwidth]{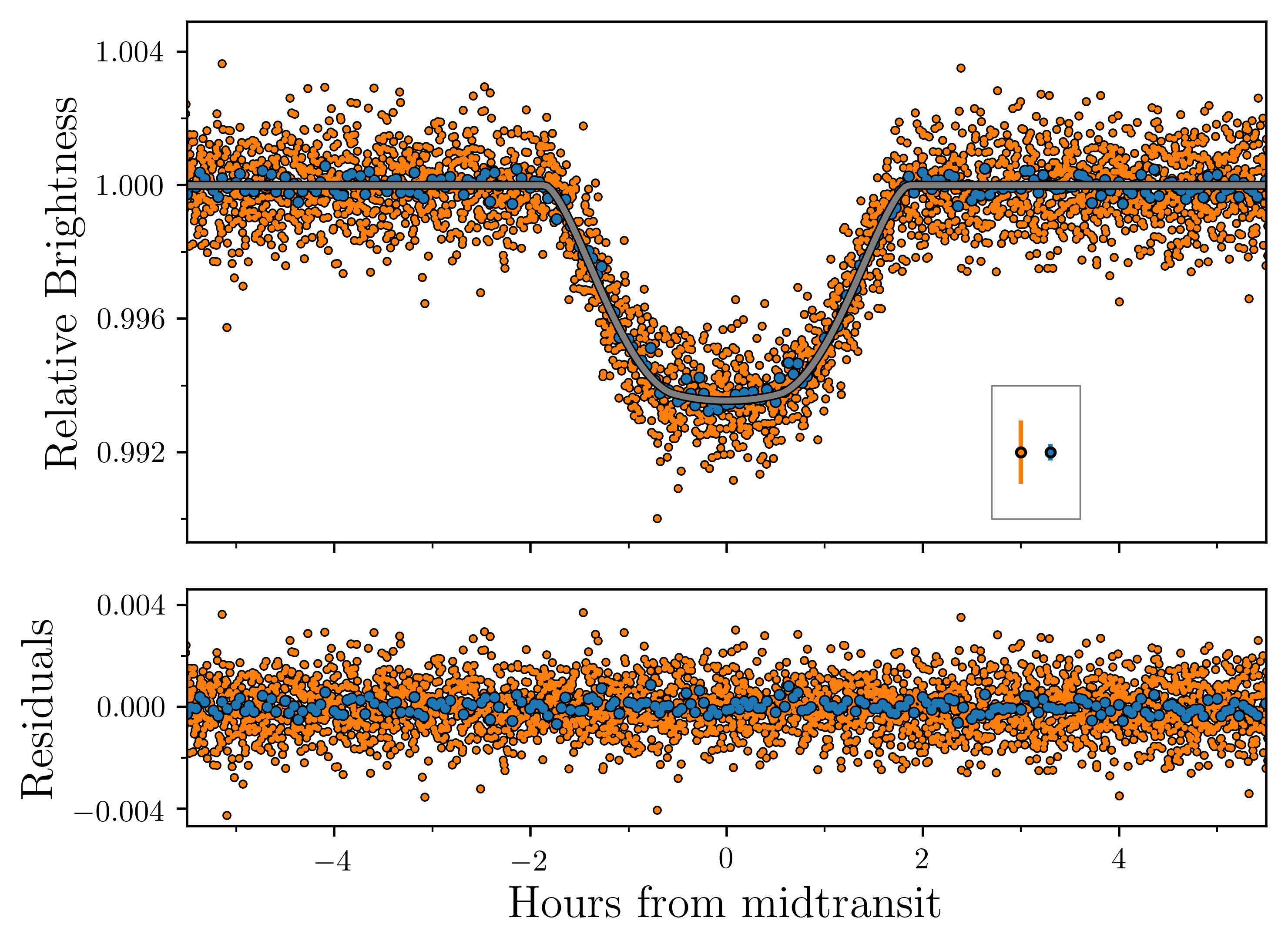}
    \caption{{\bf \tess transits of TOI-640~b.} \tess photometry phase folded to the period from \tref{tab:results} and centred on the transit. As in \fref{fig:allTESS}, blue and orange points are the 30~min and 2~min cadence data, respectively. The points with error bars in the box are not data, but illustrate the typical errors for the data. The data shown here have been detrended with the GP (white line in \fref{fig:allTESS}). The grey curve is the best-fitting light curve.}
    \label{fig:zoomTESS}
\end{figure}

In \fref{fig:zoomTESS} we show the phase-folded \tess photometry with the best-fitting transit model. We find a radius of \rpbJ for TOI-640~b. This is consistent with the value of $1.777^{+0.060}_{-0.056}$~R$_{\rm J}$ from \citet{Rodriguez2021}, but is slightly more precise owing to the additional \tess photometry and increased cadence. While we find consistent results for the radius, with \mpbJ we find a roughly $2\sigma$ difference in mass from the value reported in \citet{Rodriguez2021} of $0.88\pm0.16$~M$_{\rm J}$.

We find a value of $0.906^{+0.007}_{-0.009}$ for the impact parameter, differing by roughly 3$\sigma$ from $0.8763^{+0.0063}_{-0.0067}$ obtained by \citet{Rodriguez2021}. This then leads to a significant difference in the results for $a/R_\star$ (our $6.31^{+0.09}_{-0.07}$ versus $6.82^{+0.22}_{-0.24}$) which is correlated with the impact parameter ($b=\cos i \frac{a}{R_\star}$). These differences in the photometric solutions may be caused by \cite{Rodriguez2021} using the then available TESS photometric data from Sectors 6 and 7 together with ground-based photometry, while we have access to Sectors 6 and 7 and Sectors 33 and 34 and do not use additional ground-based data. This discrepancy might also be caused by the difference in how the light curves have been de-trended. Furthermore, it could be due to the spectroscopic transit data as the analysis of RM data can drive the result on $b$ \citep{Albrecht2022}. We investigated whether or not the results for $b$ are consistent between the different TESS observing epochs. For this, we determined $b$ only on photometric data; first on Sectors 6 and 7 only and then on Sectors 33 and 34 only, obtaining $b=0.917^{+0.006}_{-0.005}$ and $b=0.911^{+0.006}_{-0.005}$, respectively. The values are consistent with each other and our final result. We note here that given the high impact parameter for the transit of TOI-640~b, any change in orbital inclination caused by for example\ nodal precession \citep[see e.g.][]{Szabo2012,Johnson2015,Watanabe2020,Watanabe2022,Stephan2022} may be picked up by future photometric (\tess) observations. 

\begin{figure}[t]
    \centering
    \includegraphics[width=\columnwidth]{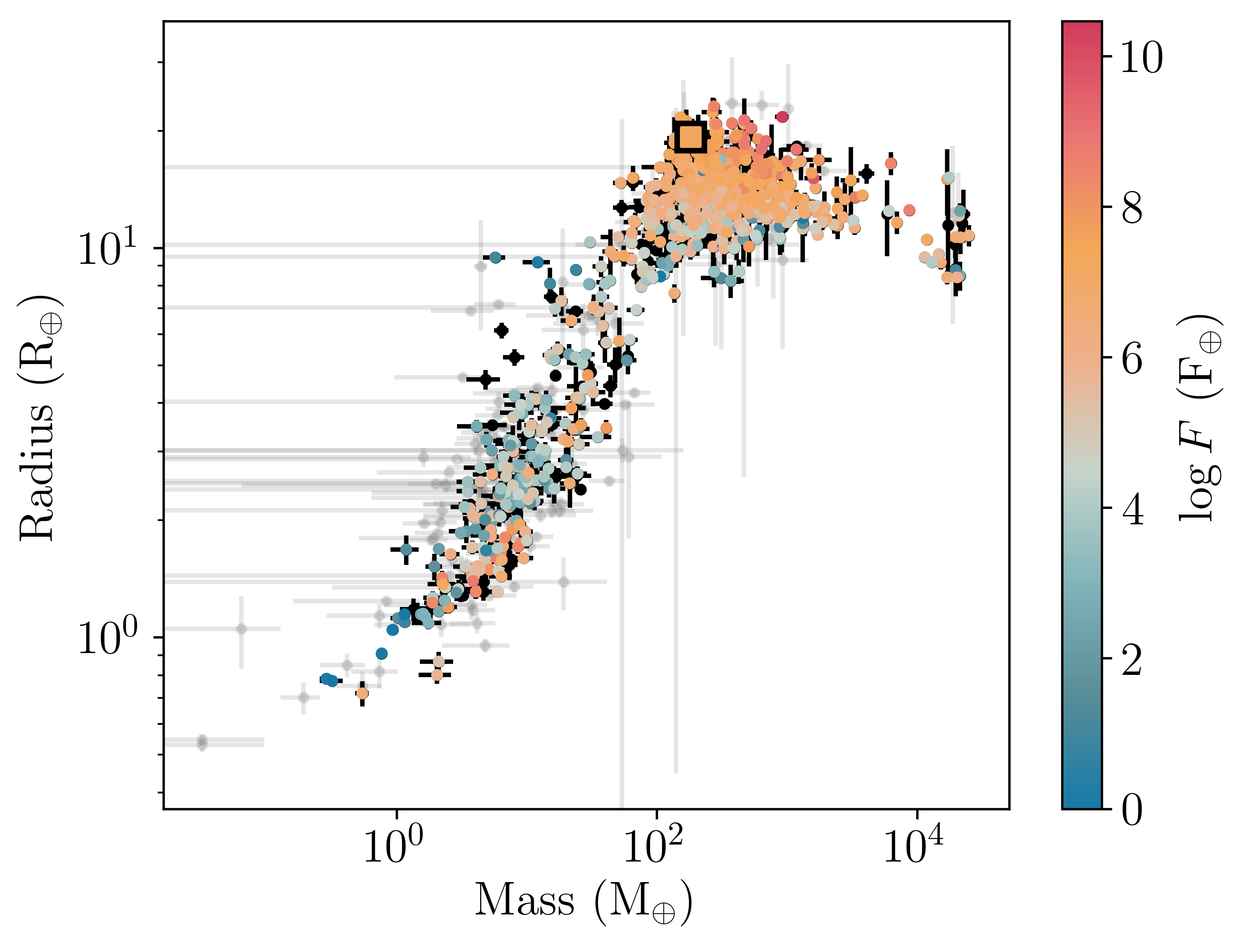}
    \caption{{\bf Mass--radius diagram.}  Here we show planets from the \texttt{TEPCat} catalogue of\ `well-studied transiting planets' \citep[as of October 2022;][ \url{https://www.astro.keele.ac.uk/jkt/tepcat/allplanets-noerr.html}]{Southworth2011}. Grey dots are measurements with uncertainties larger than 30\%, while black and coloured dots have smaller uncertainties. Points are colour coded according to the insolation, $F$, for those objects where it can be calculated. TOI-640~b is shown as the large square.}
    \label{fig:r_vs_m}
\end{figure}

As noted in \citet{Rodriguez2021}, TOI-640~b is an inflated planet. The lower mass but similar radius we find here compared to \citet{Rodriguez2021} makes it an even puffier planet with a density of \rhob. Comparing TOI-640~b to the literature, it is one of the largest and puffiest planets, but is not isolated in the mass--radius diagram as seen in \fref{fig:r_vs_m}. The puffiness is most likely due to the rather high insolation it receives.

Finally, we investigated the light curve to see if we could see any effects of gravity darkening. However, the star does not seem to be rotating fast enough to detect this effect in the \tess photometry available.

\subsection{The polar orbit of TOI-640 in context}\label{sec:polarTOI-640}

\citet{Albrecht2021} derived $\psi$ for a subset of planetary systems for which $\lambda$ measurements were available. Of the 57 systems where $\psi$ could be determined, these authors found 38 systems to be well-aligned and 18 systems misaligned in the interval between $80^\circ$ and $125^\circ$, meaning that the misaligned systems are not isotropically distributed. Rather there is a tendency for planets to be orbiting the poles of the star. 

With a value for $\psi$ of \psibrot (or \psibtworot), TOI-640 joins this preponderance of perpendicular planets. 
Given the effective temperature of the star of $6460^{+130}_{-150}$~K, which is relatively hot in this context, one might even say that our measurement of $\psi=$\psibrot is not particularly surprising, seeing as the `polar-to-aligned ratio' seems to increase with effective temperature. This might be an echo of the often larger projected obliquities found for stars with effective temperatures above $6250$~K \citep{Winn+2010}, as well as lower $v \sin i_\star$ for spectral types of F or earlier \citet{Schlaufman2010}.

\begin{figure*}
    \centering
    \includegraphics[width=\textwidth]{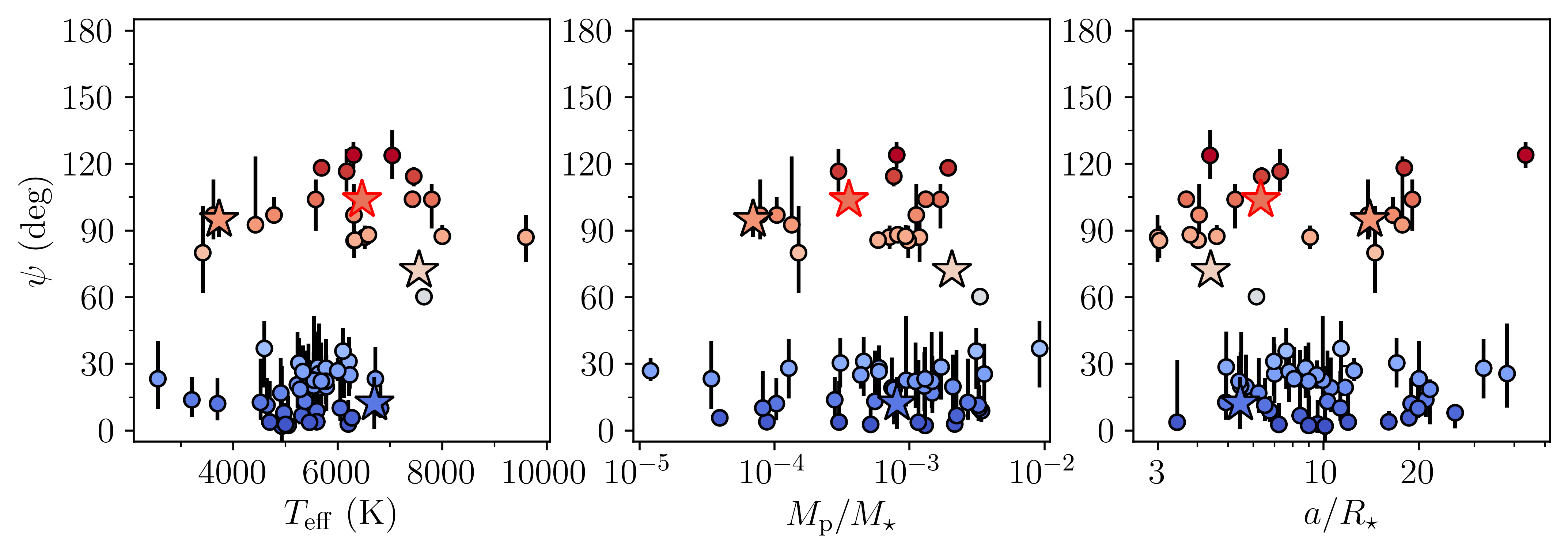}
    \caption{{\bf Preponderance of perpendicular planets.} Measurements of the 3D obliquity, $\psi$, from \citet{Albrecht2021} shown as circles with colour coding according to $\psi$. Recent $\psi$ measurements not in \citet{Albrecht2021} are shown as stars, including TOI-640 (red outline). {\it Left:} $\psi$ as a function of effective temperature. {\it Middle:} $\psi$ as a function of planet-to-star mass ratio. {\it Right:} $\psi$ as a function of orbital separation. It is worth noting that here we are only considering the results for $\psi$ corresponding to $i$ (and not $180^\circ-i$).}
    \label{fig:obls}
\end{figure*}

Further measurements of $\psi$ have recently been made. \fref{fig:obls}  shows MASCARA-1~b \citep[$\psi=72.1^{+2.5}_{-2.4}$$^\circ$;][]{Hooton2022}, GJ~3470~b \citep[$\psi=95^{+9}_{-8}$$^\circ$;][]{Stefansson2022}, KELT-7~b \citep[$\psi=12.4 \pm 11.7$$^\circ$;][]{Tabernero2022}, and TOI-640~b ($\psi=$\psibrot) along with the measurements presented in \citet{Albrecht2021}. Evidently, these new measurements follow the tendency of polar orbiting planets, where especially for hot stars harbour polar orbiting planets.

However, when discussing the sample of polar planets, it is important to keep in mind the various different selection effects that might play a role. For a classic example, see Figure~1 by \citet{Winn+2010} and for a more recent discussion on selection effects related to spin-orbit angle measurements, see \citet{Albrecht2022}. In this context, we note that we first selected TOI-640 as a system for which RM measurements, employing HARPS, are likely to result in a conclusive answer concerning $\lambda$. We only started analysing the \tess light curves to establish whether or not we were able to determine the stellar rotation period from these light curves {after} the RM measurements had been taken.

As to why planets should show a tendency to travel over stellar poles, \citet{Albrecht2021} briefly discuss four mechanisms, which we simply list here as 1) tidal dissipation \citep{Lai2012,Rogers2013,Anderson2021}, 2) Von Zeipel-Kozai-Lidov cycles \citep{Fabrycky2007}, 3) secular resonance crossing \citep{Petrovich2020}, and 4) magnetic warping \citep{Foucart2011,Lai2011,Romanova2021}. Another recent proposal was presented by \citet{Vick2022}, who highlight that a binary companion (and its influence on the obliquity during disk dissipation) combined with subsequent Kozai-Lidov cycles might also produce polar orbits. While these mechanisms might be able to explain parts of the observed distribution, they do not seem to be able to fully reproduce the observations individually. It would be interesting to increase the sample size and expand the parameter space to try to decipher whether or not these mechanisms work in tandem in different types of systems harbouring different types of planets.

\section{Conclusions}\label{sec:con}

Here we present in-transit spectroscopic measurements for the hot Jupiter TOI-640~b. These measurements allowed us to detect the RM effect both directly as the distortion of the spectral lines in the planet shadow as well as in the RVs. From this, we measured a projected spin-orbit angle for the host star of \lam. While this would suggest that the orbit of the planet is not only retrograde, but completely anti-aligned, the rotation period we measured from the \tess light curves suggests that the obliquity is \psibrot, meaning that the orbit is actually polar. 

\section{Acknowledgements}
We thank the anonymous referee for comments and suggestions which improved the manuscript.
Based on observations collected at the European Organisation for Astronomical Research in the Southern Hemisphere under ESO programme 106.21TJ.001.
We acknowledge the use of public TESS data from pipelines at the TESS Science Office and at the TESS Science Processing Operations Center. Resources supporting this work were provided by the NASA High-End Computing (HEC) Program through the NASA Advanced Supercomputing (NAS) Division at Ames Research Center for the production of the SPOC data products.
Funding for the Stellar Astrophysics Centre is provided by The Danish National Research Foundation (Grant agreement no.: DNRF106).
E.K. and S.A. acknowledge  the support from the Danish Council for Independent Research through a grant, No.2032-00230B.
The numerical  results presented in this work were obtained at the Centre for Scientific Computing, Aarhus \url{https://phys.au.dk/forskning/faciliteter/cscaa/}.
K.W.F.L. was supported by Deutsche Forschungsgemeinschaft grants RA714/14-1 within the DFG Schwerpunkt SPP 1992, Exploring the Diversity of Extrasolar Planets.
C.M.P. and M.F. gratefully acknowledge the support of the Swedish National Space Agency (DNR 65/19, 177/19, 174/18, 2020-00104).
This work is partly supported by JSPS KAKENHI Grant Number JP18H05439 and JST CREST Grant Number JPMJCR1761.
This research made use of Astropy,\footnote{http://www.astropy.org} a community-developed core Python package for Astronomy \citep{astropy2013,astropy2018,astropy:2022}. 
This research made use of matplotlib \citep{matplotlib}.
This research made use of TESScut \citep{tesscut}.
This research made use of astroplan \citep{astroplan}.
This research made use of SciPy \citep{scipy}.
This research made use of corner \citep{corner}.
This research made use of statsmodels \citep{statsmodels}.
\bibliographystyle{aa} 
\bibliography{bibliography} 

\appendix
\section{Additional tables and figures}

\begin{figure*}
    \centering
    \includegraphics[width=\textwidth]{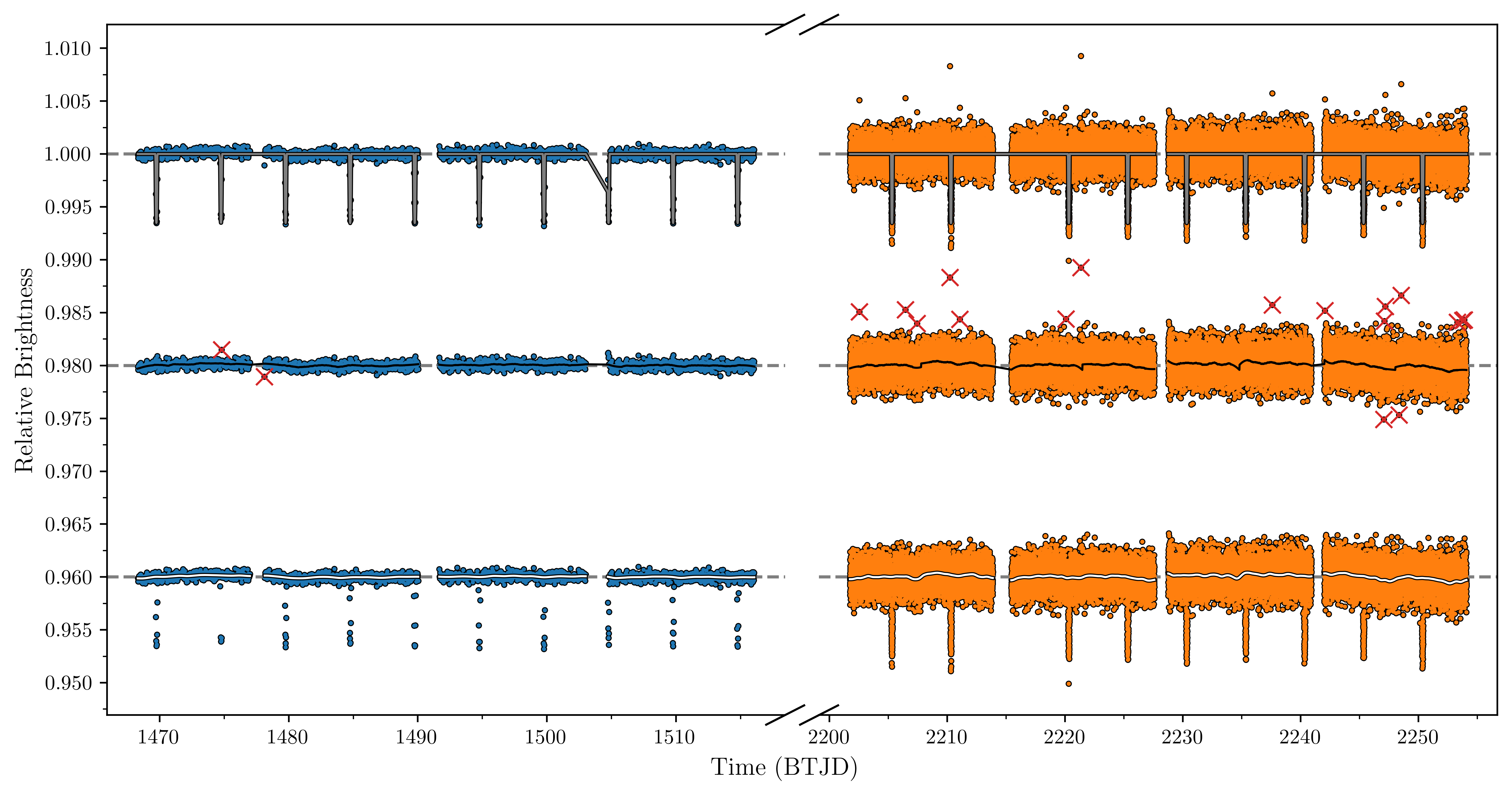}
    \caption{{\bf \tess photometry of TOI-640.} \tess photometry of TOI-640 with Sectors 6 and 7 is shown in blue to the left and Sectors 33 and 34 in orange to the right. The light curves at the  top have been corrected for scattered light. The grey curves show the best-fitting (determined iteratively) transit model. In the middle, we have subtracted this transit model. We used a Savitzky-Golay filter (black curve) to identify outliers, which are marked as red crosses. In the bottom, we have re-injected the transits into the light curves with the outliers removed. The white curves are the GPs we used to de-trend the data during our MCMC fit (see \sref{sec:RM}). }
    \label{fig:allTESS}
\end{figure*}

\begin{figure*}
    \centering
    \includegraphics[width=\textwidth]{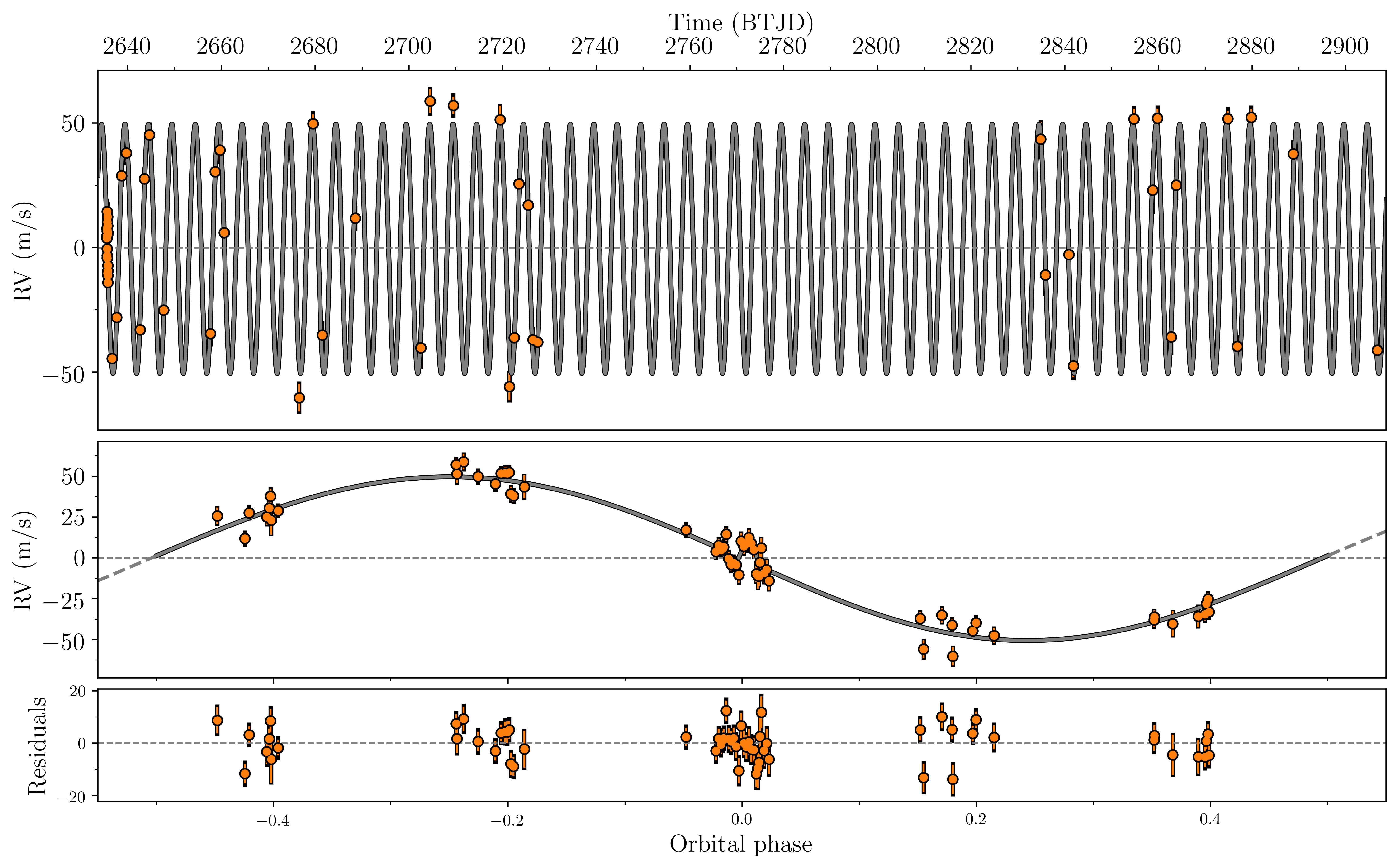}
    \caption{{\bf HARPS RVs.} {\it Top:}  HARPS RVs shown in orange with the best-fitting model overplotted in grey. {\it Middle:}  HARPS RVs phase folded to the period of the planet for the values in \tref{tab:results}. {\it Bottom:} Residuals after subtracting the Keplerian orbit and the RM effect.
    }
    \label{fig:rv_grand}
\end{figure*}

\begin{table*}
    \centering
    \caption{{\bf HARPS transit RVs.} }
    \begin{threeparttable}  
    \begin{tabular}{c c c c c}
    \toprule
    Time & RV & $\sigma$(RV) & Exp. time & SNR \\
    BJD$_{\rm TDB}$ & m~s$^{-1}$ & m~s$^{-1}$ & s & at 550~nm \\
    \midrule
2459635.51756438 & 40528.65 & 3.54 & 900 & 52.7 \\ 
2459635.52844228 & 40532.54 & 3.69 & 900 & 50.2 \\ 
2459635.53900864 & 40529.57 & 3.53 & 900 & 52.4 \\ 
2459635.54967985 & 40531.40 & 3.43 & 900 & 54.3 \\ 
2459635.56055881 & 40539.26 & 3.82 & 900 & 49.8 \\ 
2459635.57133384 & 40524.30 & 4.01 & 900 & 48.2 \\ 
2459635.58221429 & 40520.64 & 3.85 & 900 & 50.6 \\ 
2459635.59267602 & 40521.66 & 4.00 & 900 & 49.2 \\ 
2459635.60365925 & 40520.56 & 4.38 & 900 & 45.7 \\ 
2459635.61443404 & 40514.51 & 4.86 & 900 & 41.5 \\ 
2459635.62510594 & 40535.13 & 4.78 & 900 & 42.4 \\ 
2459635.63598431 & 40531.87 & 4.57 & 900 & 44.4 \\ 
2459635.64675899 & 40533.44 & 4.54 & 900 & 45.0 \\ 
2459635.65732614 & 40537.21 & 4.87 & 900 & 42.2 \\ 
2459635.66841376 & 40533.63 & 4.85 & 900 & 42.8 \\ 
2459635.67898046 & 40529.91 & 4.58 & 900 & 45.2 \\ 
2459635.68975559 & 40515.10 & 5.12 & 900 & 41.2 \\ 
2459635.70063547 & 40513.60 & 5.81 & 900 & 37.0 \\ 
2459635.71140933 & 40530.85 & 6.05 & 900 & 36.1 \\ 
2459635.72228850 & 40515.54 & 5.94 & 900 & 36.9 \\ 
2459635.73295947 & 40517.61 & 5.65 & 900 & 38.7 \\ 
2459635.74373448 & 40510.81 & 5.62 & 900 & 39.1 \\ 

    \bottomrule
    \end{tabular}
\begin{tablenotes}
    \item The time stamps, RVs and associated errors, exposure times, and S/Ns for our HARPS observations on the transit night. 
\end{tablenotes}
\end{threeparttable}  
    \label{tab:trvs}
\end{table*}

\begin{table*}
    \centering
    \caption{{\bf HARPS RVs.}  }
    \begin{threeparttable}  
    \begin{tabular}{c c c c c}
    \toprule
    Time & RV & $\sigma$(RV) & Exp. time & SNR \\
    BJD$_{\rm TDB}$ & m~s$^{-1}$ & m~s$^{-1}$ & s & at 550~nm \\
    \midrule

2459636.61429745 & 40480.25 & 2.98 & 1500 & 64.5 \\ 
2459637.61272941 & 40496.75 & 3.20 & 1500 & 60.7 \\ 
2459638.65059912 & 40553.67 & 2.90 & 1500 & 68.3 \\ 
2459639.65598973 & 40562.80 & 3.55 & 1500 & 55.8 \\ 
2459642.62712950 & 40491.86 & 3.61 & 1500 & 55.1 \\ 
2459643.53042388 & 40552.43 & 3.23 & 1500 & 58.0 \\ 
2459644.58214205 & 40570.11 & 3.68 & 1500 & 53.1 \\ 
2459647.62659055 & 40499.68 & 3.79 & 1200 & 53.5 \\ 
2459657.62127673 & 40490.31 & 3.89 & 1200 & 52.2 \\ 
2459658.62766323 & 40555.27 & 3.95 & 1200 & 51.4 \\ 
2459659.65953279 & 40564.03 & 4.29 & 1200 & 48.6 \\ 
2459660.56634930 & 40530.77 & 3.34 & 1200 & 59.4 \\ 
2459676.55929847 & 40464.58 & 5.58 & 1400 & 36.7 \\ 
2459679.53476551 & 40574.59 & 3.69 & 1400 & 53.0 \\ 
2459681.51615428 & 40489.74 & 4.41 & 1200 & 45.3 \\ 
2459688.54613223 & 40536.66 & 3.67 & 1400 & 54.9 \\ 
2459702.51767879 & 40484.57 & 7.71 & 1400 & 28.1 \\ 
2459704.49113530 & 40583.64 & 4.72 & 1200 & 43.2 \\ 
2459709.46414553 & 40581.93 & 3.51 & 1500 & 55.8 \\ 
2459719.47467143 & 40576.20 & 5.52 & 1500 & 37.6 \\ 
2459721.46803448 & 40469.00 & 5.36 & 1500 & 38.1 \\ 
2459722.45452427 & 40488.65 & 3.97 & 1500 & 50.7 \\ 
2459723.45482634 & 40550.48 & 5.03 & 1500 & 41.6 \\ 
2459725.45734693 & 40541.89 & 3.42 & 1500 & 58.5 \\ 
2459726.45743072 & 40487.81 & 3.99 & 1500 & 50.4 \\ 
2459727.45664837 & 40486.89 & 4.16 & 1500 & 48.6 \\ 
2459834.84967236 & 40568.37 & 7.04 & 1200 & 29.2 \\ 
2459835.84688757 & 40513.81 & 7.70 & 1200 & 27.6 \\ 
2459840.85990011 & 40522.06 & 9.83 & 1200 & 22.5 \\ 
2459841.86062937 & 40477.29 & 4.58 & 1200 & 42.3 \\ 
2459854.79154962 & 40576.47 & 4.13 & 1200 & 46.3 \\ 
2459858.78710518 & 40547.88 & 8.92 & 1200 & 24.2 \\ 
2459859.78571868 & 40576.71 & 3.91 & 1200 & 49.1 \\ 
2459862.74740732 & 40488.98 & 6.46 & 1200 & 32.2 \\ 
2459863.77184520 & 40549.89 & 4.76 & 1200 & 41.5 \\ 
2459874.78085079 & 40576.56 & 3.16 & 1200 & 58.0 \\ 
2459876.80989269 & 40485.16 & 3.17 & 1200 & 57.1 \\ 
2459879.81844207 & 40577.14 & 3.36 & 1200 & 54.5 \\ 
2459888.80667830 & 40562.48 & 4.45 & 1200 & 42.3 \\ 
2459906.72984965 & 40483.65 & 3.76 & 1200 & 49.3 \\ 

    \bottomrule
    \end{tabular}
\begin{tablenotes}
    \item The time stamps, RVs and associated errors, exposure times, and S/Ns for our HARPS monitoring observations carried out from 2022 February 26 to November 23.
\end{tablenotes}
\end{threeparttable}  
    \label{tab:rvs}
\end{table*}

\begin{table*}
    \centering
    \caption{{\bf Additional parameters.} }
    \begin{threeparttable}
    \begin{tabular}{c c c c c}

\toprule 
  &   &   & RV & Shadow \\ 
Parameter & Name & Prior & \multicolumn{2}{c}{Value} \\ 
\midrule 
\multicolumn{5}{c}{Stepping parameters} \\ 
\midrule 
$\log A_1$ & GP amplitude \tess 2 min. & $\mathcal{U}$ & $-8.50 \pm 0.09$ & $-8.50^{+0.09}_{-0.10}$ \\ 
$\log \tau_1$ & GP time scale \tess 2 min. ($\log$ days) & $\mathcal{U}$ & $-0.20 \pm 0.15$ & $-0.19^{+0.14}_{-0.18}$ \\ 
$\log A_2$ & GP amplitude \tess 30 min. & $\mathcal{U}$ & $-9.41^{+0.12}_{-0.13}$ & $-9.41 \pm 0.13$ \\ 
$\log \tau_2$ & GP time scale \tess 30 min. ($\log$ days) & $\mathcal{U}$ & $0.0 \pm 0.3$ & $0.0 \pm 0.4$ \\ 
$q_1 + q_2$ & Sum of limb-darkening coefficients TESS & $\mathcal{N}$(0.527,0.1) & $0.496 \pm 0.019$ & $0.496^{+0.021}_{-0.019}$ \\ 
$q_1 + q_2$ & Sum of limb-darkening coefficients HARPS & $\mathcal{N}$(0.6974,0.1) & $0.75^{+0.09}_{-0.08}$ & $0.71 \pm 0.09$ \\ 
\midrule 
\multicolumn{5}{c}{Fixed parameters} \\ 
\midrule 
$q_1 - q_2$ & Difference of limb-darkening coefficients TESS& $\mathcal{F}(-0.1028)$ & &  \\ 
 $q_1 - q_2$ & Difference of limb-darkening coefficients HARPS& $\mathcal{F}(0.2400)$ & &  \\ 
 \midrule 
\multicolumn{5}{c}{Derived parameters} \\ 
\midrule 
$q_1$ & Linear limb-darkening coefficient TESS &   & $0.197 \pm 0.010$ & $0.197^{+0.011}_{-0.010}$ \\ 
$q_1$ & Quadratic limb-darkening coefficient TESS &   & $0.300 \pm 0.010$ & $0.300^{+0.011}_{-0.010}$ \\ 
$q_1$ & Linear limb-darkening coefficient HARPS &   & $0.50 \pm 0.04$ & $0.40 \pm 0.05$ \\ 
$q_2$ & Quadratic limb-darkening coefficient HARPS &   & $0.26 \pm 0.04$ & $0.305 \pm 0.045$ \\ 
\bottomrule

    \end{tabular}
\begin{tablenotes}
    \item GP hyper parameters and limb-darkening coefficients from our MCMCs. Same as in \tref{tab:results}.
\end{tablenotes}
\end{threeparttable}    
    \label{tab:ld}
\end{table*}

\begin{figure*}
    \centering
    \includegraphics[width=\textwidth]{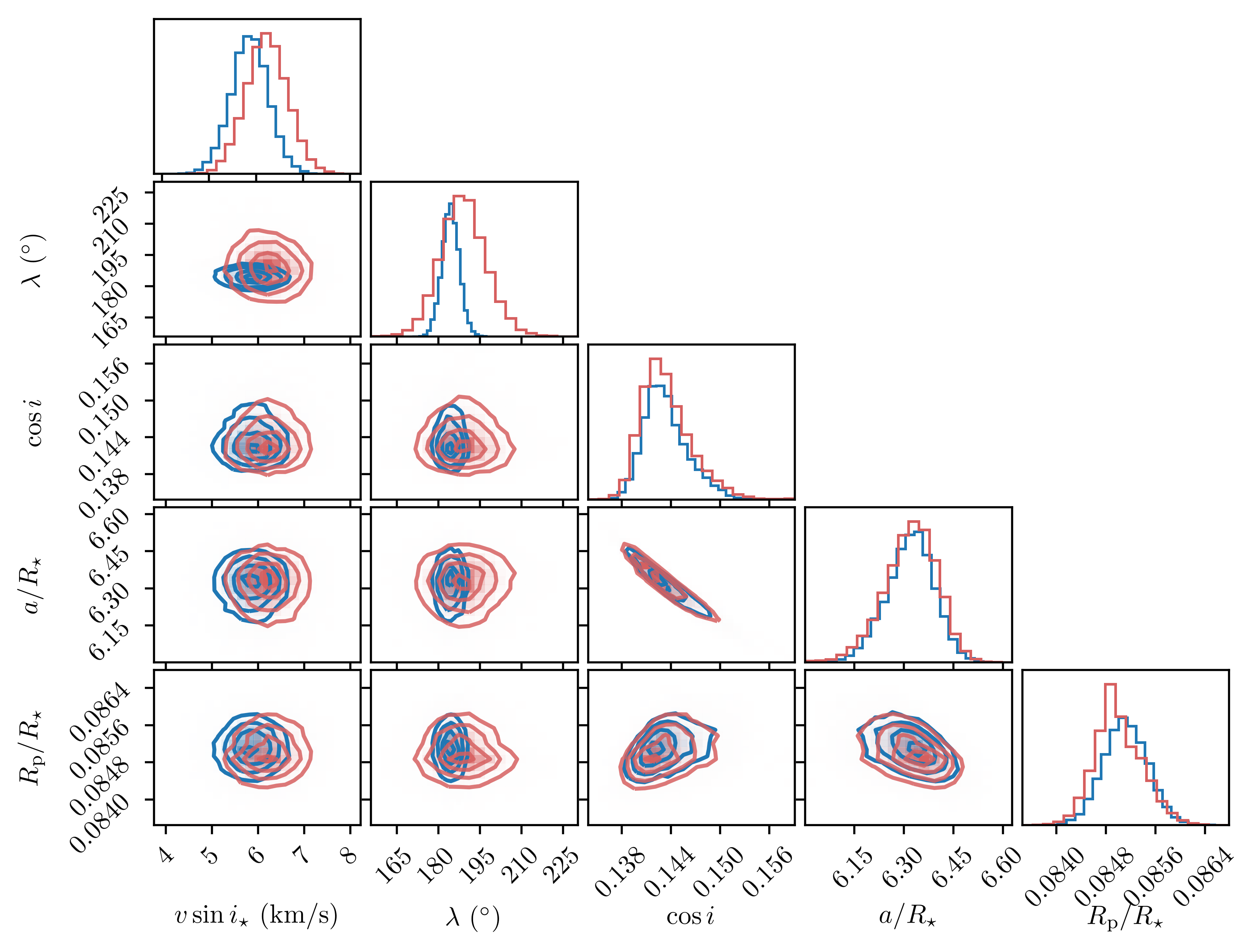}
    \caption{{\bf Correlation plots.} Here we show the correlation between some of the stepping parameters from our MCMC (\sref{sec:RM}). Blue is from our run using the RVs as input for the RM effect, and red is using the shadow.}
    \label{fig:corner_app}
\end{figure*}

\begin{figure*}
    \centering
    \includegraphics[width=\textwidth]{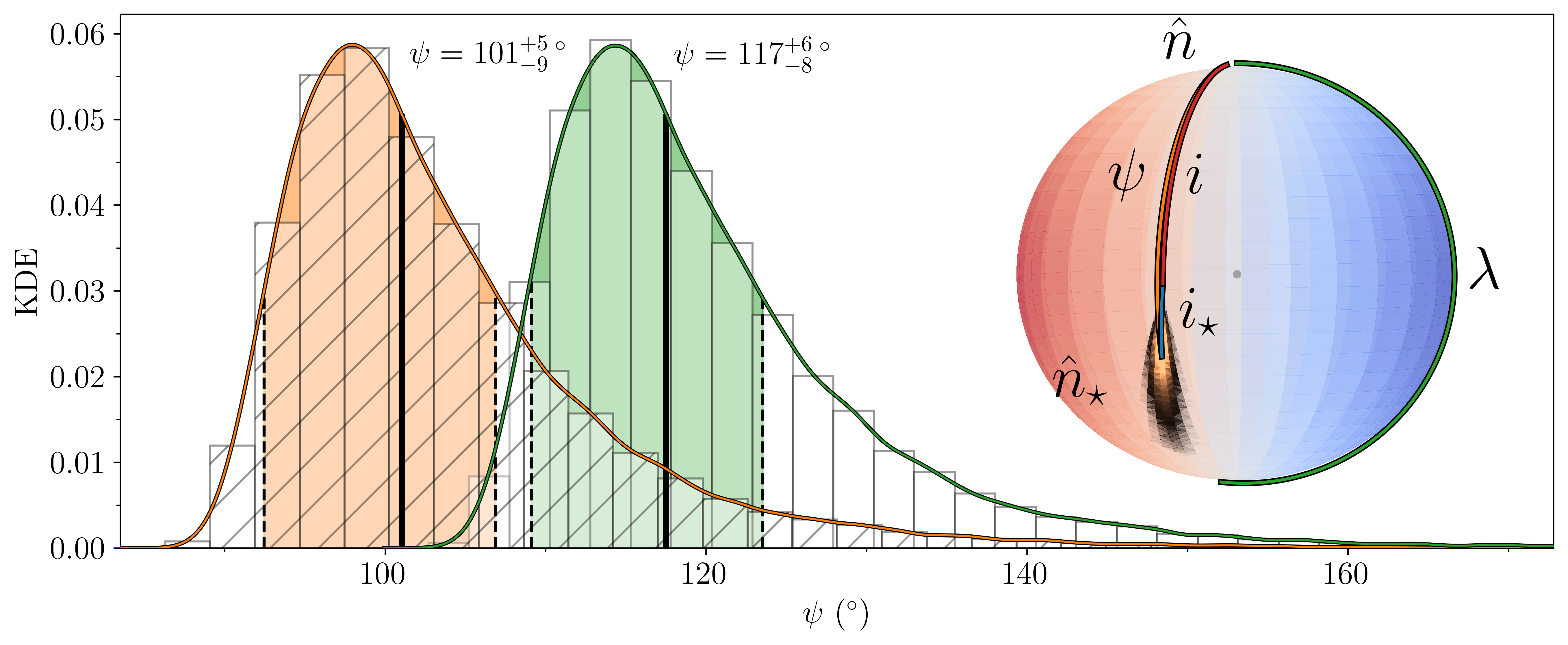}
    \caption{{\bf Spin-orbit angle distributions.} The histograms are the distributions for $\psi$ calculated from the relation in \citet{Louden2021}, otherwise the meaning is the same as in \fref{fig:psi}.
    }
    \label{fig:app_psi}
\end{figure*}

\end{document}